\documentclass{kapproc} 

\newcommand{\msun}{\mbox{M$_\odot$}}

\usepackage{graphics}

\kluwerbib

\begin{document}



\articletitle[]{Molecular gas in nearby powerful radio galaxies}


\author{S. Leon $^1$, J. Lim $^2$, F. Combes $^3$, D. Van-Trung $^2$}



\prologue{\large $^1$ University of Cologne, Germany  \\
$^2$ ASIAA, Academia Sinica, Taipei, Taiwan \\
$^3$ Observatory of Paris, France \\
}{} 

\begin{abstract}

We report the detection of $^{12}$CO($1 \rightarrow 0$) and$^{12}$CO($2 \rightarrow 1$)   emission
from the central region of nearby 3CR radio galaxies (z$<$ 0.03). Out of 21 galaxies, 8 have been
detected in, at least, one of the two CO transitions. The total molecular gas content is below 10$^9$ 
\msun. Their individual CO emission exhibit, for 5 cases, a double-horned line profile that is 
characteristic of an inclined rotating disk with a central depression at the rising part of its 
rotation curve. The inferred disk or ring distributions of the molecular gas is consistent with the 
observed presence of dust disks or rings detected optically in the cores of the galaxies. We reason
that if their gas originates from the mergers of two gas-rich disk galaxies, as has been invoked to
explain the molecular gas in other radio galaxies, then these galaxies must have merged a long time
ago (few Gyr or more) but their remnant elliptical galaxies only recently (last 10$^7$ years or less)
become active radio galaxies. Instead, we argue the the cannibalism of gas-rich galaxies provide
a simpler explanation for the origin of molecular gas in the elliptical hosts of radio galaxies
(Lim et al. 2000). 
Given the transient nature of their observed disturbances, these galaxies probably become 
active in radio soon after the accretion event when sufficient molecular gas agglomerates in their
nuclei. 

\end{abstract}


\section{Introduction}
Bright radio sources served as the first signposts for highly energetic activity in galaxies.  
The nature of these galaxies, and the reason for their luminous radio activity, have since 
been subjects of detailed investigation.  Although the vast majority resemble luminous 
elliptical galaxies, observations (Smith \& Heckman 1989) showed that a 
significant fraction of the most powerful radio galaxies at low redshifts exhibit peculiar 
optical morphologies suggestive of close encounters or 
mergers between galaxies.  Nevertheless, all of the powerful radio galaxies 
examined by Smith et al. (1990), mostly selected from the 3C catalog, lie within the fundamental 
plane of normal elliptical galaxies.  

Some radio galaxies possess so much dust that they can be detected in the far-infrared by 
{\it IRAS}.  This discovery has motivated many subsequent searches for 
molecular gas in radio galaxies, in all cases by selecting those with known appreciable 
amounts of dust.  At low redshifts, eleven radio galaxies have so far been detected in 
$^{12}$CO.  The detected galaxies 
span nearly three orders of magnitude in radio luminosity, and comprise those exhibiting 
core-dominated radio sources as well as classical double-lobed FR-I (edge darkened) and 
FR-II (edge brightened) radio sources.  All have inferred molecular-gas masses between 
$10^{9} {\rm \ M_{\odot}}$ and $10^{11} {\rm \ M_{\odot}}$, except for the very nearby 
radio galaxy Centaurus~A which has a molecular-gas mass of $\sim$$2 \times 10^{8} 
{\rm \ M_{\odot}}$ (Eckart et al. 1990).  In four of the five cases the CO gas is found to 
be concentrated 
in a compact (diameter of a few kpc or smaller) ring or disk around the center of the 
galaxy. The molecular gas in radio galaxies may therefore comprise the reservoir 
for fueling their central supermassive black holes.

At low redshifts, the only other type of galaxy to commonly exhibit molecular gas masses 
as high as $\ge 10^{10} {\rm \ M_{\odot}}$ are infrared-luminous galaxies.  These gas and 
dust rich galaxies exhibit vigorous star formation thought to be triggered by 
galaxy-galaxy interactions, and indeed the majority of ultraluminous infrared galaxies 
(i.e., those with $L(60{\mu}{\rm m}) \ge 10^{12} L_{\odot}$) are found to be merging 
systems of gas-rich disk galaxies.  In the latter systems, the CO gas is found to be 
preferentially concentrated in a ring or disk around the nuclear regions of the merging 
galaxies (e.g Bryant \& Scoville 1999).  The observed similarities have prompted the suggestion 
that radio galaxies also originate from the mergers of two gas-rich disk galaxies, and 
in many cases comprise their still disturbed E/SO merger products (Mirabel et al. 1989, Evans
et al. 1999a, 1999b).

Given the predisposition of the abovementioned surveys towards relatively 
dust-rich objects, are the radio galaxies so far detected biased towards those with unusually 
large amounts of molecular gas?  It is not clear whether 
the IR-bright radio galaxies so far detected represent the most gas-rich members of a 
population that all possess substantial amounts of dust and gas, or extreme members of 
a population that possess a broad range of dust and gas masses.  This issue is of importance 
for a proper understanding of the nature of radio galaxies, and what fuels their 
supermassive black holes.

To address this issue, we have initiated a deep survey of all the previously undetected 
radio galaxies at redshifts $z \le 0.031$ in the revised 3C catalog (Spinrad et al 1985).  
The objects in this catalog represent the most luminous radio galaxies  at their particular redshifts 
in the northern hemisphere.  At low redshifts the vast majority can be clearly 
seen to be luminous elliptical galaxies, which in most cases comprise first ranked galaxies 
in poor clusters of roughly ten members although a number reside in much richer 
environments (Zirbel 1997).  Only a small minority have 
been detected in the infrared by {\it IRAS}; indeed, only three objects in this catalog 
have previously been detected in CO, one of which lies in the redshift range of our survey.

\section{Molecular gas}

\begin{figure}
\center{\resizebox{10cm}{8cm}{\includegraphics{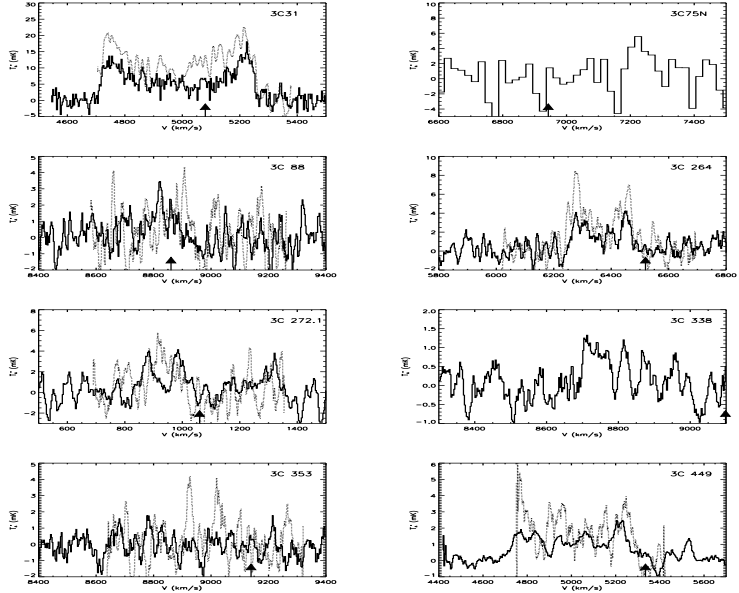}}} 
\caption{CO(1-0) spectra (solid line) with CO(2-1) spectra (dotted line) towards a sample
of 3CR radio galaxies detected in, at least, one of the both transition lines. The vertical
arrows show  the optical velocity.}
\label{spectra}
\end{figure}

Out of the  21 sources observed, 8 have been detected in $^{12}$CO($1 \rightarrow 0$) and 3 in 
$^{12}$CO($2 \rightarrow 1$) (see Table \ref{tab_co}). The line profiles of the  
$^{12}$CO($1 \rightarrow 0$) and $^{12}$CO($2 \rightarrow 1$) emission are shown in
Fig. \ref{spectra}. Because of the half-power beam of the IRAM-30m, we are mainly sensitive
to emission from the central part of the radio-galaxies, typically the inner 10/5 kpc for
the ($1 \rightarrow 0$)/($2 \rightarrow 1$) transitions. We emphasize nevertheless
that the CO beam sizes are larger than the dust features seen in these radio-galaxies (Martel et al. 
1999).

The double-horned line profile is  clearly observed in 5 radio-galaxies and is characteristic of an
inclined rotating disk with a central depression in CO emission at the rising part of its rotation 
curve (Wiklind et al. 1997). The rotational velocities observed are quite high and if 
they are corrected for the inclination
of the dust disk they reach high values which are more typical of the nuclear molecular-gas disks or
rings in IR-luminous galaxies. Using a standard CO-to-H$_2$  conversion factor
we estimate the molecular content (see Table 1) within these
galaxies. In the case of 3C253, where we have only a detection in $^{12}$CO ($2 \rightarrow 1$,
we compute the molecular mass using the mean average line ratio.

The line ratio of $^{12}$CO($2 \rightarrow 1$/$^{12}$CO($1 \rightarrow 0$) , with the 
individual line 
intensities measured provides information on the opacity and excitation of the gas. Line ratio less 
than 0.7 imply that the gas has different filling factor in the two transitions or is subthermally 
excited, whereas line ratio greater than unity imply that the gas is optically thin. If both 
transitions originitate from a region comparable in size with their individual dust features, as is 
the case in all the galaxies so far mapped in CO, the line ratios are between 0.6 and 0.8. These values
are close to the extreme upper limits measured for many inactive elliptical galaxies (Wiklind et al. 
1995),
but close to the average value of 0.9 measure at the centers of both inactive and active disk galaxies
(Braine \& Combes 1992).

\begin{table}
\begin{tabular}{lcccc}
Name & Velocity & Transition & I(CO) & M(H$_2§$)  \\
     &  (km.s$^{-1}$) &       & (K.km.s$^{-1}$) & (Log(\msun)) \\
\hline
3C31    &  5071  &  $^{12}$CO ($1 \rightarrow 0$) & 3.87    &  9.02 \\
        &        &  $^{12}$CO ($2 \rightarrow 1$) & 4.80    &       \\
3C75N   &  6816  &  $^{12}$CO ($1 \rightarrow 0$)  & 0.37    &  8.26 \\
3C88    &  8859  &  $^{12}$CO ($1 \rightarrow 0$)  & 0.18    &  8.19 \\
3C264   &  6523  &  $^{12}$CO ($1 \rightarrow 0$)  & 0.45    &  8.30 \\
        &        &  $^{12}$CO ($2 \rightarrow 1$)  & 0.93    &       \\ 
3C272.1 &  1060  &  $^{12}$CO ($1 \rightarrow 0$)  & 0.35    &  6.61 \\
        &        &  $^{12}$CO ($2 \rightarrow 1$)  & 0.77    &       \\
3C338   &  9100  &  $^{12}$CO ($1 \rightarrow 0$)  & 0.10    & 7.94  \\
3C353   &  9150  &  $^{12}$CO ($2 \rightarrow 1$)  & 0.20    & 7.98  \\
3C449   &  5345  &  $^{12}$CO ($1 \rightarrow 0$)  & 0.86    & 8.37  \\
\end{tabular}
\caption{Radio-galaxies detected in  $^{12}$CO transitions in our survey.}
\label{tab_co}
\end{table}

\section{Discussion}
\begin{figure}
\resizebox{6cm}{5cm}{\includegraphics{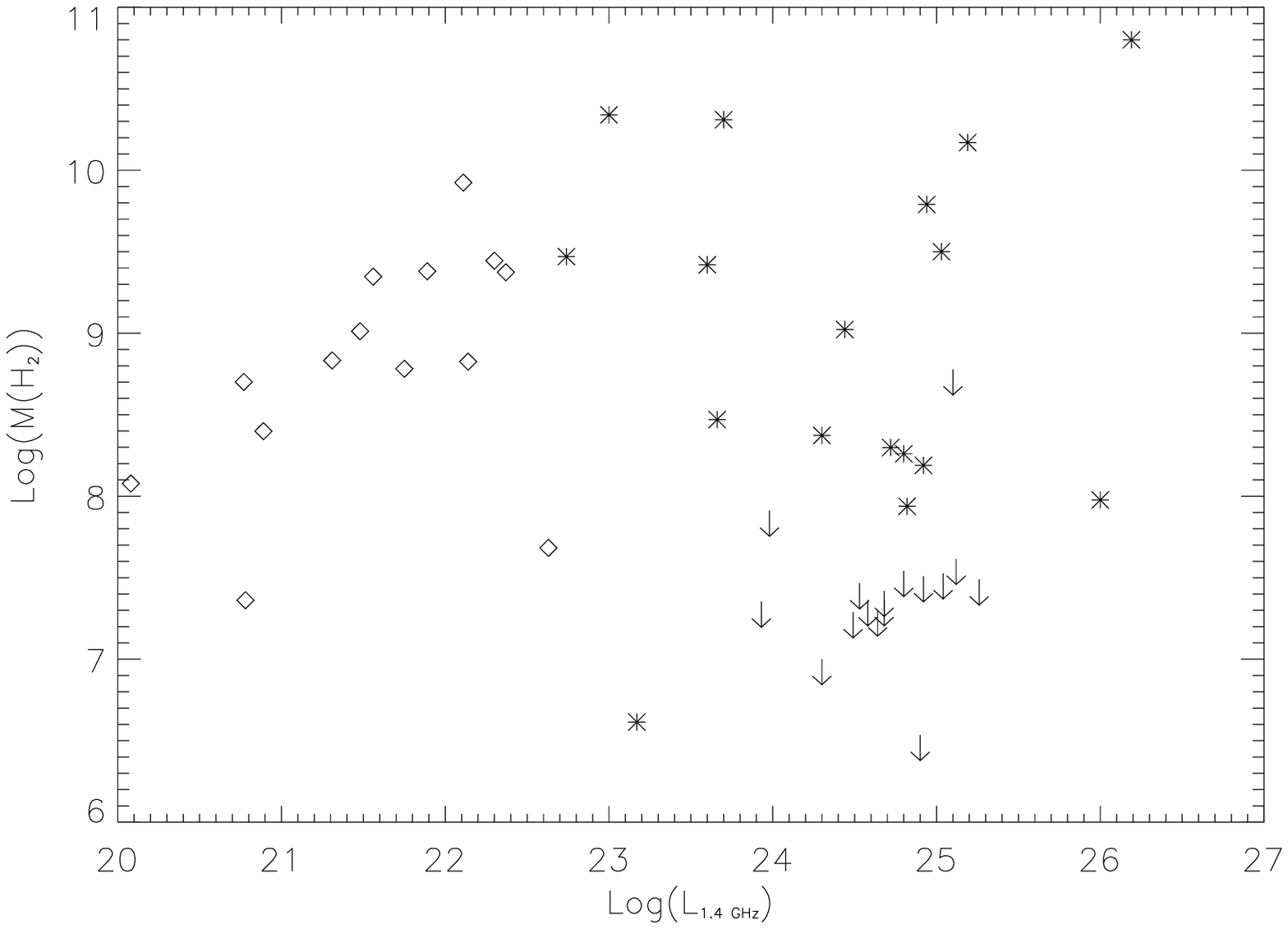}}
\resizebox{6cm}{5cm}{\includegraphics{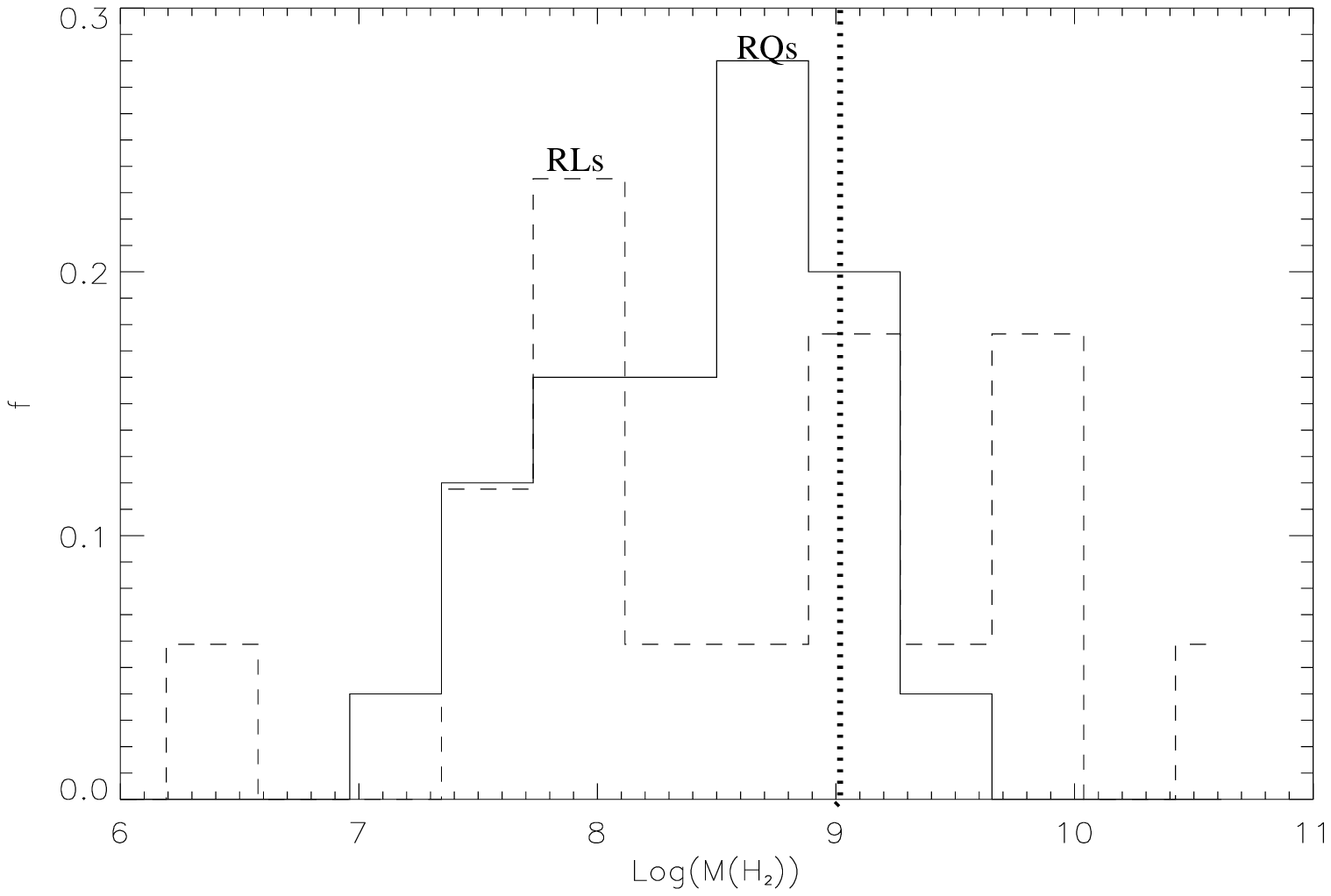}}
\caption{Left: Molecular gas content vs. radiocontinuum (1.4 GHz). Right: 
Histogram of the molecular content in the radio-galaxies (RGs) and elliptical radio-quiet. The vertical
dash line is at the maximum molecular gas mass detected in our 3CR sample.
galaxies  (RQs).}
\label{frequency}
\end{figure}

The total/upper-limit molecular mass found in this 3CR sample of radio galaxies is well below the 
molecular mass found in a typical galaxy like the Milky Way (several 10$^9$ \msun) for
 most of the case.
We detected only $4\times 10^6$ \msun molecular gas in the nearby radio-galaxy 3C272.1 (M84) and an 
upper-limit of $3.4\times 10^6$ in 3C270 (M87). These low values contrast with the previous 
high-content molecular gas found mainly in  IRAS-selected  radio-galaxies (Mazzarrela et al. 1993, 
Evans et al. 1999a,b). On the Fig. \ref{frequency} the distribution of the 
molecular gas mass in the radio galaxies 
show the dichotomy bettween the IRAS-selected sample of radio-galaxies  and our 3C sample: a clear 
cut appears at $10^9 \msun$ between both samples. Compared to a sample of radio-quiet elliptical 
galaxies (Knapp \& Rupen 1996, Wiklind et al. 1997) the 3C sample exhibits a statiscally 
significant lower gas  mass content than the elliptical galaxy sample.

A possible link between Ultra-Luminous Infrared Galaxies (ULIRGs) and radio-galaxies has been
proposed (Evans et al. 1999b). Some case exhibits properties that place them in both
categories. The question is if major mergers are responsible for the AGN phase in radio galaxies.
By comparison with ULIRGs, however, radio galaxies exhibit a much broader range of molecular
 gas masses.
Moreover the host galaxies of the 3C sample appear to be very well relaxed and lie in the fundamental
plane. The timescale for relaxation after a merging is about 1-2 Gy. Nevertheless the radio emission
phase is {\it very short}, few $10^7$ years. Given the accretion rate for the AGN 
($< 1$ \msun.yr$^{-1}$), only few massive GMCs ($10^6 \msun$) would be sufficient to fuel the AGN.
Indeed in the nearby radio galaxies (M84, M87) observed, the detection or upper-limit of
molecular gas  are lower than $10^7$ \msun. In our survey only low molecular gas content has been
detected. Major mergers require radio galaxies to be very old merger remnants that have only recently
become active after the remnant has relaxed and much of the gas disappeared. But minor mergers 
present a simpler alternative to the dust and molecular gas seen in many 3CR radio galaxies. 
Furthermore they can  explain the presence of a molecular disk and the loss of angular 
momentum necessary to bring the gas towards the center.

%




\end{document}